\documentstyle[eqsecnum,aps,prd]{revtex} 

\begin{document}
\draft
\twocolumn[\hsize\textwidth\columnwidth\hsize\csname
@twocolumnfalse\endcsname
\title{\bf
Complete constraints on a nonminimally coupled chaotic 
inflationary scenario from the cosmic microwave background
}

\author{
Eiichiro Komatsu\footnotemark[1]
and Toshifumi Futamase\footnotemark[2]
}

\address{\small\sl
Astronomical Institute, Graduate School of Science, 
Tohoku University, Sendai 980-8578, Japan
}                      

\maketitle

\begin{abstract}
\indent
We present {\em complete} constraints imposed from observations of
the cosmic microwave background radiation (CMBR) on the chaotic
inflationary scenario with a nonminimally coupled inflaton
field proposed by Fakir and Unruh (FU).
Our constraints are complete in the sense that we investigate both 
the scalar density perturbation and the tensor gravitational wave in
the Jordan frame, as well as in the Einstein frame.
This makes the constraints extremely strong
without any ambiguities due to the choice of frames.
We find that the FU
scenario generates tiny tensor contributions to the CMBR relative to
chaotic models in minimal coupling theory, in spite of
its spectral index of scalar perturbation being slightly tilted.
This means that the FU scenario
will be excluded if any tensor contributions to CMBR are detected
by the forthcoming satellite missions. Conversely, if no tensor nature 
is detected despite the tilted spectrum, a minimal chaotic scenario
will be hard to explain and the FU scenario will be supported.

\end{abstract}


\vskip2pc]

\footnotetext[1]{Email address: komatsu@astr.tohoku.ac.jp}
\footnotetext[2]{Email address: tof@astr.tohoku.ac.jp}

\section{Introduction}


\indent
In spite of its many successes, the standard big-bang theory 
has faced serious problems, namely, the horizon, flatness, 
and monopole problems. 
In the beginning of the 1980s, an epoch-making idea called 
the inflationary scenario was advocated to solve these cosmological
puzzles\cite{Sato,Guth}.
Later it was recognized that the concept gives us not only 
a solution to such puzzles,
but also to the origin of density 
perturbations\cite{GP,Linde82,Hawking,Starobinsky}.


Among the various models of the inflationary scenario, 
Linde's chaotic model\cite{Linde83} has been regarded as
a feasible and natural mechanism for the realization of inflationary
expansion. This model still has a serious problem; i.e., one has to fine-tune 
the self-coupling constant $\lambda$ of the inflaton unacceptably
small to have a reasonable amplitude
of the density perturbations.


On the other hand, the feasibility of inflation has been investigated 
in alternative theories of gravity, e.g., the Brans-Dicke scalar tensor 
theory\cite{LS89,SA90}, and nonminimal coupling theories of
gravity\cite{FM89,AZT85}. Fakir and Unruh (FU) \cite{FU90,FU90b} proposed
a way to avoid fine-tuning $\lambda$ by introducing a relatively
large nonminimal coupling constant $|\xi|>1$ in the context of 
the chaotic inflationary model. 
According to their results, the large value of $\xi$,
i.e., order of $10^3$, allows us to have a reasonable value for 
the coupling constant $\lambda=10^{-2}$.
Thus the FU scenario remains a reasonable model of the
inflationary scenario.


Constraints on the FU scenario are discussed by some
authors\cite{Salopek92,Kaiser95} using the scalar perturbations 
generated during the inflationary phase. 
We investigated the spectrum of tensor mode cosmic microwave
background radiation (CMBR) anisotropy
\cite{KF98}. 
Hwang also discussed the tensor mode power spectrum 
from inflation based on generalized gravity theories in a unified
manner\cite{Hwang98}.


In discussing the constraint on generalized gravity theories
including the FU scenario, one has to be careful about ambiguities associated with 
the conformal transformation. Sometimes the analysis is made in the
conformally transformed frame in which the gravity may be described by
the Einstein action. However, it has long been known that 
the conformal transformation often changes the physical
phenomena in different frames (e.g., Ref. \cite{Dick98},
and references therein).  
Thus it is quite important to make 
the frame dependences of the results one obtains clear. 

The purpose of the present paper is to investigate the constraints on
the FU scenario by taking into account the frame dependency. 
Namely, we shall investigate the CMBR anisotropy caused 
by both the scalar and tensor perturbations in two different
frames, the Jordan and Einstein frames, which seem to have special
physical importance among various transformed frames. 
From this point of view, this work can be regarded as the {\em complete} 
treatment of the observational constraints on the FU scenario.

\begin{table}[h]
\caption{
Predicted parameters based on the Fakir-Unruh scenario.
For comparison, the parameters of a minimal coupling chaotic
scenario are also shown.
All parameters are derived at $N(t_k)=70$.
}
\label{sum}
\begin{center} 
  \begin{tabular}{ccc}
  Parameter       & Fakir-Unruh scenario & minimal chaotic scenario \\
  \hline 
  $r$             & $2\times 10^{-3}$         & 0.2 \\
  $n_s$     & 0.97                      & 0.96 \\
  $n_t$     & $-3.0\times 10^{-4}$      & $-2.8\times 10^{-2}$ \\
  $\lambda/\xi^2$ & $4\times 10^{-10}$        & \\
  \end{tabular}
\end{center}
\end{table}


A number of investigations have shown that high precision CMBR
temperature anisotropy and polarization
experiments, e.g., two satellite missions of NASA's
Microwave Anisotropy Probe (MAP)\cite{MAP}
and ESA's Planck Surveyor\cite{PLANCK}, can be used to determine
many cosmological parameters to
unprecedented precision\cite{Jung96,BET97,ZSS97}. 
We are especially interested in amplitudes and spectral indices of 
scalar and tensor perturbations. 
It is relatively hard to determine these
parameters due to the cosmic variance and the cosmic
confusion\cite{Knox95,Bond94,EB98} by means of a temperature spectrum
only. Including polarization informations allows us to detect tensor
contributions directly because a tensor mode can generate the magnetic
mode of polarization while a scalar mode cannot
\cite{HW97,SZ97,Ztheis}.
However, it is
still hard to detect such a magnetic mode directly because of
its predicted tiny amplitude.
Even Planck, with the most sensitive experiment not only for temperature
anisotropy but also for polarization, can detect
tensor contributions only if the tensor to scalar ratio is greater
than 0.2\cite{SZ97,Ztheis}.

  
For later convenience we summarize the relevant facts of the 
CMBR experiments here. The experiments can measure the angular power 
spectrum of the temperature
or polarization correlation function $C_l$,
\begin{equation}
  \frac{\delta T(\theta,\phi)}{T_0}
  = \sum_{l,m}a_{l m}Y_{l m}(\theta,\phi),
\end{equation}
\begin{equation}
  \left<a_{l' m'}^*a_{l m}\right>
  = C_l\delta_{l'l}\delta_{m'm},
\end{equation}
where the angle brackets denote ensemble average.
The Cosmic Background Explorer (COBE) Differential Microwave
Radiometer (DMR) group expressed the observed quadrupole moment in terms
of $Q_{\rm rms-PS}$,
\begin{equation}
  Q_{\rm rms-PS}
  \equiv T_0\sqrt{\frac{5 C_2}{4\pi}}.
\end{equation}
According to the COBE four-year results\cite{Fixsen,Bennett}, 
\begin{eqnarray}
  T_0 &=& 2.728\pm 0.004 \ {\rm K}\label{cobe1},\\
  Q_{\rm rms-PS} &=& 18\pm 1.4\ {\rm \mu K}\label{cobe2}
\end{eqnarray}
for the Harrison-Zel'dovich spectrum. This gives $C_2^{\rm obs}=1.1\times
10^{-10}\ {\rm \mu K^2}$.


This paper is organized  as follows. In Sec. II we review the 
background solutions of the inflationary expansion in both 
the Jordan and Einstein frames. 
In Secs. III and IV, we show the amplitude of the scalar curvature 
perturbation and the tensor gravitaional wave generated during the
de Sitter phase, and discuss 
the constraints by means of the observed CMBR quadrupole moment. 
Section V derives the predicted tensor to scalar ratio and
describes the possibility of detecting tensor contributions. 
To compare our results with well-known results in minimal coupling theory 
and to interpret their physical meanings, we also discuss  
a {\em consistency relation} which includes the spectral index in Sec. VI. 
Finally, Sec. VII contains conclusions. 
Table \ref{sum} summarizes our results of the predicted observables
based on the FU scenario.
We shall follow 
Misner, Thorne, and Wheeler\cite{MTW} for the definition of 
the Riemann tensor, Ricci tensor, and Ricci scalar, but the metric convention
is chosen as $\mbox{\boldmath{$g$}}=(+$$-$$-$$-)$.

\section{Background inflationary solutions}

\subsection{Jordan frame solutions}

At first, we shall review the background inflationary solutions in the 
original Jordan frame.
This section follows our previous paper\cite{KF98}.
We shall consider the following action:
\begin{equation}
  \label{Jaction}
  A = \int{d^4x \sqrt{-g}
           \left[
	         \frac R{2\kappa^2}
		 +\frac12\xi\phi^2R
		 -\frac12g^{\mu\nu}\phi_{,\mu}\phi_{,\nu}
		 +V(\phi)
	   \right]
	  },
\end{equation}
where $\kappa^2\equiv 8\pi G$. 
Our definition of $\xi$ is the same as Fakir and Unruh\cite{FU90}, that is,
conformal coupling means $\xi = -1/6$.
Note that Futamase and Maeda\cite{FM89} used an opposite sign for $\xi$.
For the spatially flat Robertson-Walker metric
\begin{equation}
  ds^2
  = dt^2 - a^2(t)\delta_{ij}dx^idx^j,
\end{equation}
we can derive the fundamental background equations
\begin{equation}
  H^2=\frac{\kappa^2}{3(1+\kappa^2\xi\phi^2)}
      \left[
            \frac12{\dot{\phi}}^2
	    +V(\phi)
	    -6\xi H\phi\dot{\phi}
      \right],
\end{equation}
\begin{eqnarray}
  \nonumber
  \ddot{\phi}&
  +&3H\dot{\phi}
  +\left[
         \frac{\kappa^2\xi\phi^2(1+6\xi)}{1+\kappa^2\xi\phi^2(1+6\xi)}
   \right]\frac{\dot{\phi}^2}{\phi}\\
  \nonumber
  \mbox{}&=& \frac1{1+\kappa^2\xi\phi^2(1+6\xi)}
             \left[
	           4\kappa^2\xi\phi V(\phi)
		   -(1+\kappa^2\xi\phi^2)V_{,\phi}
	     \right],\\	     
\end{eqnarray}
where overdots denote time derivatives in the Jordan frame and 
$V_{,\phi} \equiv {\partial V}/{\partial \phi}$. 
Now we shall employ the potential
$V(\phi)=\lambda\phi^4/4$ and apply ordinary slow-roll
approximations to the background equations.
This gives us
\begin{eqnarray}
  \label{backeq}
  H^2
  &\approx& \frac{\kappa^2\lambda\phi^4}{12(1+\kappa^2\xi\phi^2)}
            \left[
	          1+\frac{8\xi}{1+\kappa^2\xi\phi^2(1+6\xi)}
            \right],\\
  \label{backphi}    
  3H\dot{\phi}
  &\approx& -\frac{\lambda\phi^3}{1+\kappa^2\xi\phi^2(1+6\xi)}.
\end{eqnarray}
It is straightforward to find the self-consistent inflationary
solutions under the condition $\kappa^2\xi\phi^2 \gg 1$.
Defining $\psi\equiv \kappa^2\xi\phi^2$, the above
equations take the following simple forms:
\begin{equation}
  H^2 = \frac{\lambda\psi}{12\kappa^2\xi^2},\qquad
  \frac{\dot{\psi}}{H} = -\frac{8\xi}{1+6\xi}.
\end{equation}
These solutions lead to the well-known exponential expansion in the
Jordan frame. The amount of expansion from any epoch to the end of
inflation is calculated as
\begin{equation}
  \label{N}
  N(t) \equiv \int_{t}^{t_f}{
                             H dt
			    }
       = \int_{\psi(t)}^{\psi_f}{
                                 \frac{H}{\dot{\psi}} d\psi
				}
       = \frac{1+6\xi}{8\xi}\left[\psi(t)-\psi_f\right].
\end{equation}
Note that for the initial value of $\psi$,
\begin{equation}
  N(t_i) = \frac{1+6\xi}{8\xi}(\psi_i-\psi_f)
         \approx \frac{1+6\xi}{8\xi}\psi_i\ge 70
\end{equation}
must be held to solve the cosmological puzzles.

\subsection{Einstein frame solutions}

We shall perform the conformal transformation to the Einstein frame
\begin{equation}
  \hat{g}_{\mu\nu} = \Omega g_{\mu\nu},\qquad
  \Omega = 1 + \kappa^2\xi\phi^2.
\end{equation}
Hereafter we put hats on variables defined in the Einstein frame. 
The conformal transformation gives
\begin{equation}
  \label{Eaction}
  A = \int{d^4x\sqrt{-\hat{g}}
           \left[
	         \frac{\hat{R}}{2\kappa^2}
                 -\frac12F^2(\phi)\hat{g}^{\mu\nu}\phi_{,\mu}\phi_{,\nu}
                 +\hat{V}(\phi)\
	   \right]
	  },
\end{equation}
where
\begin{equation}
  F^2(\phi)
  \equiv \frac{1+\kappa^2\xi\phi^2(1+6\xi)}{(1+\kappa^2\xi\phi^2)^2}
\end{equation}
and
\begin{equation}
  \label{potent}
  \hat{V}(\phi)
  \equiv \frac{V(\phi)}{(1+\kappa^2\xi\phi^2)^2}
  = \frac{\lambda \phi^4}{4(1+\kappa^2\xi\phi^2)^2}
  \approx \frac{\lambda}{4\kappa^4\xi^2},
\end{equation}
where the last equality of Eq. (\ref{potent}) is derived from the
condition $\kappa^2\xi\phi^2\gg 1$.

To make the kinetic term of scalar field canonical form, we redefine 
the scalar field as
\begin{equation}
  \label{Ptrans}
  \frac{d\hat{\phi}}{d\phi}
  = F(\phi)
  = \frac{\sqrt{1+\kappa^2\xi\phi^2(1+6\xi)}}{1+\kappa^2\xi\phi^2}.
\end{equation}
Then it can be clearly seen that the new potential
(\ref{potent}) is still
flat enough to lead to sufficient exponential inflation. 
When we investigate the dynamics of the universe in the Einstein
frame, we should transform our coordinate system to make the metric 
the Robertson-Walker form
\begin{equation}
  \hat{a}
  = \sqrt{\Omega}a,\qquad d\hat{t}
  = \sqrt{\Omega}dt,
\end{equation}
and we obtain
\begin{equation}
  d{\hat s}^2
  = d{\hat t}^2
    - \hat{a}^2(\hat{t})\delta_{ij}dx^idx^j.
\end{equation}
Note that the physical quantities in the Einstein frame should be defined in 
this coordinate system. Now the Einstein equation can be derived
in the usual manner under the slow-roll approximations,
\begin{equation}
  \hat{H}^2
  = \frac{\kappa^2}3 \left[
                           \left(
			         \frac{d\hat{\phi}}{d\hat{t}}
	                   \right)^2
			   + \hat{V}(\hat{\phi})
		     \right]
  \approx \frac{\lambda}{12\kappa^2\xi^2},
\end{equation}
where
\begin{eqnarray}
  \label{Htrans}
  \hat{H}
  &\equiv& \frac1{\hat{a}}\frac{d{\hat{a}}}{d\hat{t}}
  = \frac1{\sqrt{\Omega}} \left(
                                H + \frac12\frac{\dot{\Omega}}{\Omega}
			  \right),\\
  \label{Pdottrans}		  
  \frac{d\hat{\phi}}{d\hat{t}}
  &=& \left(
            \frac{d\hat{\phi}}{d\phi}
      \right) \left(
                    \frac{dt}{d\hat{t}}
	      \right) \dot{\phi}
  = \frac{\sqrt{1+\kappa^2\xi\phi^2(1+6\xi)}}{\Omega^{3/2}}\dot{\phi}.
\end{eqnarray}

We can conclude that the exponential behavior of the expansion 
is retained in both frames\cite{FM89}. Note that we can put a constraint on
$\lambda/\xi^2$ by means of requiring $\hat{V}< m_{\rm pl}^4$,
\begin{equation}
  \frac{\lambda}{\xi^2}
  < 256\pi^2.
\end{equation}
This constraint is too weak compared with
the observational constraints discussed below.

\section{Scalar perturbation}

The scalar curvature perturbation
${\cal{R}}(t,\mbox{\boldmath$x$})$
generated as the quantum noise during
the de Sitter phase is well-known in the Einstein
frame. Let us choose the longitudinal gauge
\begin{equation}
  ds^2 = \left[
               1+2\Psi(x)
	 \right] dt^2
	 - a^2(t) \left[
	                1 + 2\Phi(x)
	          \right] \delta_{ij}dx^idx^j.
\end{equation}
We can construct the gauge-invariant scalar curvature perturbation from the metric and
the inflaton field perturbation as follows:
\begin{equation}
  {\cal R}(x) = \Phi(x) - \frac{H}{\dot{\phi}}\delta\phi(x).
\end{equation}
Since we already have the prescriptions to quantize $\Phi$ and
$\delta\phi$ in the Einstein frame\cite{MS91,MFB92}, we can calculate the 
amplitude of scalar curvature perturbation
\begin{eqnarray}
  \sqrt{\hat{P}_S(k)}
  &\equiv& \sqrt{\frac{4\pi k^3}{(2\pi)^3}
           \int{d^3\mbox{\boldmath$x$}
	        e^{i\mbox{\boldmath$k$}\cdot\mbox{\boldmath$x$}}
		\left<
		      \hat{\cal{R}}(\hat{t},\mbox{\boldmath$0$})
		      \hat{\cal{R}}(\hat{t},\mbox{\boldmath$x$})
		\right>}}\\
  &=& \left.
            \frac{\hat{H}^2}{2\pi|d\hat{\phi}/d{\hat{t}}|}
      \right|_{\hat{t}_k},
\end{eqnarray}
where $k$ is a comoving wave number.
Note that the metric perturbations defined in the Einstein frame
have to be calculated in the coordinate system \{$\hat{x}^{\mu}$\} :
\begin{equation}
  \hat{g}_{\mu\nu}(\hat{x})
  = \frac{\partial x^{\alpha}}{\partial \hat{x}^{\mu}}
    \frac{\partial x^{\beta}}{\partial \hat{x}^{\nu}}
    \hat{g}_{\alpha\beta}(x)
\end{equation}
and for instance,
\begin{equation}
  \hat{\Phi}(\hat{x})
  = \Phi(x) + \frac12\frac{\delta\Omega}{\Omega}(x).
\end{equation}
Makino and Sasaki\cite{MS91} and Fakir, Habib, and Unruh\cite{FHU92}
proved that the amplitude of scalar perturbation in the Jordan frame
exactly coincides with that in the Einstein frame.
We can see such conformal invariance in the simple calculation
\begin{eqnarray}
  \hat{\cal{R}}(\hat{x})
  &=& \hat{\Phi}(\hat{x})
      - \frac{\hat{H}}{d\hat{{\phi}}/d\hat{t}}\delta\hat{\phi}(\hat{x}),\\
  &=& \Phi(x)
      - \frac{H}{\dot{\phi}}\delta\phi(x)
  = {\cal R}(x).
\end{eqnarray}
This proof allows us to calculate the
scalar power spectrum in the Jordan frame quite easily,
\begin{equation}
  \label{S}
  \sqrt{P_S(k)}
  = \frac1{2\pi\sqrt{1+6\xi}}\left.
                                   \frac{H^2}{|\dot{\phi}|}
			     \right|_{t_k}
  = \frac{N(t_k)}{2\pi}\sqrt{\frac{\lambda}{3\xi(1+6\xi)}},
\end{equation}
where we used slow-roll approximations, and we have a corrected
missing factor of 2 in the original paper of Makino 
and Sasaki\cite{MS91}. The curvature perturbation
gives the
Newtonian potential perturbation $\Psi$\cite{MS91,S93},
\begin{eqnarray}
  \sqrt{\left<
              \left|
	            \Psi
	      \right|^2
	\right>_k}
  &=& \frac23\sqrt{P_S(k)}
      \qquad \mbox{radiation-dominated era},\\
  &=& \frac35\sqrt{P_S(k)}
      \qquad \mbox{matter-dominated era}.
\end{eqnarray} 
Since the observed CMBR quadrupole anisotropy is dominated by the
Sachs-Wolfe (SW) effect\cite{SW67,HS95}, we can simply estimate it as
\begin{eqnarray}
  \nonumber
  C_2^{\rm scalar} &=& \left<
                       \left(
		             \frac{\delta T}{T_0}
		       \right)^2
		 \right>_{\rm SW}
	       = \left<
	               \left|
		             \frac13\Psi
		       \right|^2
		 \right>_{k=d_H^{-1}}\\
  \label{C2PS} 		 
               &=& \frac1{25}P_S(k=d_H^{-1}),
\end{eqnarray}
where $d_H$ is the present Hubble horizon scale.
We can thus constrain the set of parameters
\begin{equation}
  \frac{\lambda}{\xi^2} < 4.0\times 10^{-10},
\end{equation}
where the inequality takes into account the contribution from a tensor
mode. If we adopt $\lambda=10^{-2}$, it gives $\xi > 
5\times 10^3$. 

\section{Tensor perturbation}
For completion, let us show our previous result\cite{KF98}
of the amplitude of tensor perturbation in the Jordan frame.
In the synchronous gauge, the metric becomes
\begin{equation}
  ds^2 = a^2(\tau)
               \left[
	             d\tau^2
                      - \left(
	                      \delta_{ij}
		              + h_{ij}
		        \right)dx^idx^j
	       \right],
\end{equation}
where $\tau$ is a conformal time and $h_{ij}$ is a
transverse-traceless perturbation.
Note that $h_{ij}$ is invariant under any conformal
transformations\cite{MFB92}.
The power spectrum of
the tensor perturbation can be derived as well as the scalar one, 
\begin{eqnarray}
  \nonumber
  \sqrt{\hat{P}_T(k)}&=&\sqrt{P_T(k)}\\
  \nonumber
  &\equiv& \sqrt{
                 \frac{4\pi k^3}{(2\pi)^3}2
                 \sum_{\lambda=+,\times}
		 \int{d^3\mbox{\boldmath$x$}
	                     e^{i\mbox{\boldmath$k$}\cdot\mbox{\boldmath$x$}}
		             \left<
		                     h_{\lambda}(\tau,\mbox{\boldmath$0$})
                                     h_{\lambda}(\tau,\mbox{\boldmath$x$})
		             \right>
		            }
	        }\\
  \label{T}	   
  &=& \frac{4}{\sqrt{\pi}m_{\rm pl}}\frac H{\sqrt{1+\psi}}
  \approx \frac1{\pi}\sqrt{\frac{\lambda}{6\xi^2}},
\end{eqnarray}
where $\lambda=+,\times$ are modes of the polarization.

Since a tensor perturbation also causes temperature anisotropy via
SW effect as well as a scalar one, we can constrain
$\lambda/\xi^2$ from the
CMBR observations\cite{KF98,White92}. 
\begin{equation}
  \label{C2PT}
  C_2^{\rm tensor}\simeq 0.0363P_T
\end{equation}
gives
\begin{equation}
  \frac{\lambda}{\xi^2} < 1.8\times 10^{-7},
\end{equation}
where the inequality also takes into account the contribution
from the scalar
mode. $\lambda=10^{-2}$ gives $\xi > 2\times 10^2$. 

\section{Constraints from the cosmic microwave background}

Now we are in a position to predict the ratio of the amplitude of
the tensor perturbation to that of the scalar one.
It can be obtained from Eqs. (\ref{S}) and (\ref{T}), 
\begin{equation}
  \label{mainresult}
  \frac{P_T}{P_S}(k) = 0.00245 \frac{1+6\xi}{6\xi}
                               \left(
                                     \frac{70}{N(t_k)}
			       \right)^2. 
\end{equation}
We should keep in mind that this result does not depend on the choice
of frames and does not depend on $\xi$ directly in the limit of $\xi\gg 
1$ but depends on $N(t_k)$ only.
Equations (\ref{C2PS}) and (\ref{C2PT})
give the simple relation between $C_2^{\rm tensor}/C_2^{\rm scalar}$ and $P_T/P_S$,
\begin{equation}
  \label{r}
  r
  \equiv \frac{C^{\rm tensor}_2}{C^{\rm scalar}_2}
  \simeq 0.9 \frac{P_T}{P_S}(k=d_H^{-1}).
\end{equation}
We already know that the FU scenario requires $\xi \gg 1$ to avoid
the fine-tuning of $\lambda$, and the perturbations which contribute
to the present CMBR quadrupole moment have left the Hubble horizon scale
at around $N(t_k)=70$.
We can thus conclude that the FU scenario predicts $r\simeq 0.002$,
which is too tiny to be detected by even the Planck Surveyor. In
other words, if MAP or Planck could detect any tensor
contributions to the CMBR, it would mean that the FU scenario 
could be excluded from good candidates of the inflationary model.

\section{Spectral Indices}

In the previous sections, we discussed only the amplitudes of
perturbations. Here let us
consider the first-order solutions in the slow-roll approximations. The
behavior of perturbations are fully governed by the simple Schr\"odinger
type equation, and the first-order nature appears in the
time-dependent mass term $R''/R$,
\begin{equation}
  (R\Delta)''(\mbox{\boldmath$k$},\tau)
  + \left(k^2-\frac{R''}R\right)(R\Delta)(\mbox{\boldmath$k$},\tau)
  = 0,
\end{equation}
where $\Delta$ is a scalar or tensor perturbation\cite{Hwang96},
and dashes denote conformal time derivatives.
Writing $R=a\sqrt{Q}$, 
the usual slow-roll parameter $\epsilon$ \cite{SL93,LPB94}
and a new parameter $\alpha$ can be defined as
\footnote{
Correspondences to literatures are 
$\hat{\alpha}_s=\epsilon+\delta$\cite{SL93},
$\hat{\alpha}_s=2\epsilon-\eta$\cite{Kaiser95,LPB94},
and $\hat{\alpha}_s=-\delta/2$\cite{Bond94,CBDES93}.
}
\begin{equation}
  \epsilon\equiv -\frac{\dot{H}}{H^2},
  \qquad \alpha \equiv \frac{\dot{Q}}{2HQ}.
\end{equation}
$\epsilon$ and $\alpha$ give rise to the spectral index of the scalar
and the tensor mode 
\begin{eqnarray}
  \label{ns}
  n_s
  &\equiv& 1 + \frac{d\ln{P_S}}{d\ln{k}}
  = 1 - 2\epsilon - 2\alpha_s,\\
  \label{nt}
  n_t
  &\equiv& \frac{d\ln{P_T}}{d\ln{k}}
  = -2\epsilon - 2\alpha_t,
\end{eqnarray}
and the consistency relation
\footnote{
It is widely known that there
is a specific relation between $r$, $n_s$ and $n_t$.
Such a relation is called a consistency relation.
We should emphasize that $\alpha_s \neq \alpha_t$ in general and
the widely used simplest relation $n_t = n_s - 1$
must not be universal. 
}
\begin{equation}
  n_t = n_s - 1 + 2(\alpha_s - \alpha_t).
\end{equation}
All we should do is calculate $\Delta$ and $Q$ for both the scalar and
tensor modes in both frames.

\subsection{Scalar perturbation}

Hwang calculated $\Delta$ and $Q$ for scalar perturbation in the Jordan
frame directly\cite{Hwang97},
\begin{equation}
  \label{Qs}
  \Delta_S
  = {\cal R}_{\delta\phi=0},
  \qquad Q_S = \frac{\dot{\phi}^2
               + (3/2) (\dot{\Omega}^2/\Omega)}
                 {\left(
		        H + (1/2) (\dot{\Omega}/\Omega)
		  \right)^2},
\end{equation}
where ${\cal R}_{\delta\phi=0}$ is the scalar curvature perturbation 
in the uniform scalar field gauge, i.e., $\delta\phi=0$ and
$\Omega=1+\psi$ is the conformal factor defined previously. 

In the Einstein frame, we can use the well-known results from the
minimal coupling theory
\begin{equation}
  \hat{\Delta}_S
  = \hat{{\cal R}}_{\delta\hat{\phi}=0},
    \qquad \hat{Q}_S = \left(
                             \frac{d\hat{\phi}/d\hat{t}}{\hat{H}}
		       \right)^2.
\end{equation}
$\hat{Q}_S$ is conformally transformed as
$\hat{Q}_S = Q_S/{\Omega}$,
and it gives $\hat{a}\sqrt{\hat{Q}_S}=a\sqrt{Q_S}$,
i.e., $\hat{R}=R$.
Defining another slow-roll parameter which appears in Eq. (\ref{Qs}) as
\begin{equation}
  \beta \equiv \frac{\dot{\Omega}}{2H\Omega},
\end{equation}
we can conformally transform the slow-roll parameters
\begin{eqnarray}
  \hat{\epsilon}
  &=& \frac{\epsilon+\beta}{1+\beta}
      - \frac{\dot{\beta}}{H(1+\beta)^2}
  \sim \epsilon + \beta,\\
  \hat{\alpha}_s
  &=& \frac{\alpha_s-\beta}{1+\beta}
  \sim \alpha_s - \beta,
\end{eqnarray}
and the spectral index is also transformed as
\begin{equation}
  \hat{n}_s
  = 1 - 2\hat{\epsilon} - 2\hat{\alpha}_s
  = 1 - 2\epsilon - 2\alpha_s
  = n_s.
\end{equation}
Thus, we can conclude that $n_s$ is invariant under the conformal
transformation up to the first-order of the slow-roll approximations.  

Now we are in a position to calculate $n_s$
explicitly, 
\begin{equation}
  \hat{\epsilon}
  \sim \frac1{2\kappa^2} \left(
                               \frac{\hat{V}_{,\hat{\phi}}}{\hat{V}}
			 \right)^2
  = 1.5 \times 10^{-4}\frac{1+6\xi}{6\xi} \left(
                                                \frac{70}{N(\hat{t}_k)}
				          \right)^2,
\end{equation}
\begin{eqnarray}
  \nonumber
  \hat{\alpha}_s
  &\sim& \frac1{\kappa^2}\left[
                             \left(
			           \frac{\hat{V}_{,\hat{\phi}}}
				        {\hat{V}}
			     \right)^2
			     - \left(
			             \frac{\hat{V}_{,\hat{\phi}\hat{\phi}}}
				          {\hat{V}}
			       \right)
		       \right] \\
  \nonumber
  &=& 1.4 \times 10^{-2} \left(
                             \frac{70}{N(\hat{t}_k)}
		       \right)
    + 3.0 \times 10^{-4} \frac{1+6\xi}{6\xi} \left(
                                                   \frac{70}{N(\hat{t}_k)}
				             \right)^2.\\
\end{eqnarray}
With these quantities, we can rewrite the amplitude of perturbations as
\begin{equation}
  \sqrt{\hat{P}_S(k)}
  = \frac{\hat{H}}{\sqrt{\pi}m_{\rm pl}}
      \left.
            \frac1{\sqrt{\hat{\epsilon}}}
      \right|_{\hat{t}_k},
  \qquad
  \sqrt{\hat{P}_T(k)}
  = \left.
            \frac{4\hat{H}}{\sqrt{\pi}m_{\rm pl}}
      \right|_{\hat{t}_k},
\end{equation}
and Eq. (\ref{r}) gives 
\begin{equation}
  \label{TS}
  \hat{r}
  \simeq 14\hat{\epsilon}
  = 7 (1-\hat{n}_s-2\hat{\alpha}_s).
\end{equation}
Note that $\hat{r}$ depends on the potential steepness 
$\hat{\epsilon}$ only.

We find that slow-roll parameters do not depend on $\xi$ directly in 
the limit of $\xi\gg 1$ but depend on $N(t_k)$ as well as $r$.
Here note that $O(\hat{\epsilon})\sim O(\hat{\alpha}^2_s)$.
Although it seems to be
inconsistent with the first-order analysis, any higher-order terms
than first-order are not important here, so it is sufficient for our
purposes.

Finally, the spectral index of the scalar curvature
perturbation can be calculated :
\begin{equation}
  \label{FUparam}
  n_s = 0.97,
  \qquad
  r = \hat{r} = 0.002
\end{equation}
at $N(\hat{t}_k)=70$.
Although $n_s$ is slightly tilted,
the predicted $r$ is still too small.
Let us refer to a consistency relation here.
We can see from
Eq. (\ref{TS}) that the simplest relation $\hat{r}=7(1-\hat{n}_s)$ is held 
only if $2\hat{\alpha}_s/(1-\hat{n}_s) \ll 1$, but Eq. (\ref{FUparam})
shows $2\hat{\alpha}_s/(1-\hat{n}_s) \sim 1$.
Therefore, we must not use such a simple relation
as that widely used to analyze CMBR power spectrum
(e.g., Refs. \cite{Bond94,CBDES93}).

It is worth comparing the above results with the well-known
results in a minimal chaotic scenario.
Employing $V_m(\phi_m)=\lambda\phi_m^4/4$, we obtain
\begin{eqnarray}
  \epsilon_m
  &=& \frac8{\kappa^2 \phi_m^2}
  = 1.4 \times 10^{-2} \left(
                             \frac{70}{N(t_m{}_{,k})}
		       \right),\\
  \alpha_m{}_{,\rm s}
  &=& \frac4{\kappa^2 \phi_m^2}
  = 0.7 \times 10^{-2} \left(
                             \frac{70}{N(t_m{}_{,k})}
		       \right),
\end{eqnarray}
where $N(t_m)=\kappa^2 \phi_m^2(t_m)/8$, and
$N(t_m{}_{,k})=70$ gives
\begin{equation}
  n_m{}_{,\rm s} = 0.96,
  \qquad
  r_m = 0.2.
\end{equation}
Since $2\hat{\alpha}_m{}_{,s}/(1-\hat{n}_m{}_{,\rm s})
\sim 0.3$, we 
still should not use the simplest relation.
These results are very interesting. While both the FU and
minimal chaotic scenarios give similar tilted spectra, the amount of 
the tensor contributions to the CMBR is quite different.
This is because of
the difference of the order of $\epsilon$ between each of these theories.
Physically, the scalar field in the FU scenario moves {\em
much slower} than in the minimal one. It can be found in the flatness 
of the potential
\begin{eqnarray}
  \hat{V}_{\rm FU}
  &=& \frac{\lambda\phi^4}{4(1+\kappa^2\xi\phi^4)^2}
  \approx \frac{\lambda}{4\kappa^4\xi^2},\\
  V_m
  &=&\frac{\lambda}4\phi_m^4.
\end{eqnarray}
Equation (\ref{ns}) shows that the tilted spectrum is produced by both the
steepness and curvature of the potential shape, but Eq. (\ref{TS}) shows that
the tensor to scalar ratio is determined by the steepness only. This
is why the tensor contributions to the CMBR are quite
different between each of the theories.
As a result, we can determine which theory governs our universe by
means of the observation of CMBR temperature anisotropy and
polarization.   

\subsection{Tensor perturbation}

We have already derived $\Delta$ and $Q$ for tensor perturbation
in the Jordan frame\cite{KF98},
\begin{equation}
  \Delta_T = h_{\lambda},
                       \qquad Q_T = \Omega,
\end{equation}
and in the Einstein frame
\begin{equation}
  \hat{\Delta}_T = \hat{h}_{\lambda},
                             \qquad {\hat Q}_T = 1.
\end{equation}
We thus find
${\hat Q}_T = Q_T/{\Omega}$
and $\hat{\alpha}_t = 0 = \alpha_t - \beta$.
The spectral index of the tensor mode can be calculated in both frames,
\begin{equation}
  \hat{n}_t
  = -2\hat{\epsilon}
  = -2\epsilon - 2\beta
  = n_t
  \simeq -3.0 \times 10^{-4}.
\end{equation}
$n_t$ is also conformally invariant and we can see that
the tensor perturbation is almost scale invariant in 
the FU scenario. 
There is another expression of the consistency relation
\begin{equation}
  r = \hat{r}
    \simeq -7\hat{n}_t
    \simeq 0.002.
\end{equation}
%

\section{Conclusions}

We have investigated the feasibility of the FU scenario,
which is the chaotic
inflationary scenario characterized by a large value of the nonminimal
coupling constant, by means of the forthcoming CMBR experiments. 
We have calculated the ratio of the quadrupole contribution of 
the tensor mode to one of the scalar mode.
As a result, if any experiment could
detect the tensor gravitational wave contributions to the CMBR under
current sensitivities, the FU scenario would be
excluded from good candidates of the inflationary model. In addition, we 
discussed the spectral index of the scalar perturbation to 
make sure of the consistency of our results. Even if the spectral index is
tilted by $n_s=0.97$, the tensor contributions are still too
small to be detected. This is derived from the flatness of the potential
slope in the FU scenario. However, if no evident 
tensor contributions were detected despite the tilted spectrum,
a minimal chaotic scenario would fail and the FU scenario would be
more plausible.
Table \ref{sum} shows the summary of our results.

We found 
that the physical observables $r$, $n_s$, and $n_t$ do not depend on
$\xi$ in the limit of $\xi\gg 1$ but depend on $N(t_k)$ only.
It should be emphasized that all of
them do not depend on the choice of frames, that is,
they are {\em conformally invariant}, so 
our results can be compared to observations directly without
any ambiguities.

All of the results derived here are valid in both the Jordan 
and Einstein frames, and include both the scalar and 
tensor contributions one up to the first-order
in the slow-roll approximations.
In this sense, this work could be stated as {\em complete}.

\section*{Acknowledgments}

We would like to thank J. Hwang for fruitful discussions.



\end{document}